\documentclass[aps,pre,twocolumn,floatfix,showpacs]{revtex4}
\usepackage{epsf,graphicx}
\begin{document}
\title{Waiting time dynamics of priority-queue networks}
\author{Byungjoon~Min}
\affiliation{Department of Physics, Korea University, Seoul 136-713, Korea}
\author{K.-I.~Goh}
\thanks{Email: kgoh@korea.ac.kr}
\affiliation{Department of Physics, Korea University, Seoul 136-713, Korea}
\author{I.-M.~Kim}
\affiliation{Department of Physics, Korea University, Seoul 136-713, Korea}
\date{\today}
\begin{abstract}
We study the dynamics of priority-queue networks, generalizations
of the binary interacting priority queue model introduced by
Oliveira and Vazquez [Physica A {\bf 388}, 187 (2009)]. 
We found that the original AND-type protocol for interacting
tasks is not scalable for the queue networks with loops
because the dynamics becomes frozen due to the priority conflicts. 
We then consider a scalable interaction protocol, an OR-type one,
and examine the effects of
the network topology and the number of queues on the waiting
time distributions of the priority-queue networks,
finding that they exhibit power-law tails in all cases
considered, yet with model-dependent power-law exponents.
We also show that the synchronicity in task executions, 
giving rise to priority conflicts in the priority-queue networks,
is a relevant factor in the queue dynamics that can change
the power-law exponent of the waiting time distribution. 
\end{abstract}
\maketitle

In the last century, queueing theory has proved useful for various problems 
ranging from operations research to telecommunications \cite{queueing}.
There is a recent resurgence of interest for the queueing theory 
among the statistical physics community
with the application to the problems in human dynamics.  
Specifically, various queueing models based on the prioritization of tasks,
or the priority queue models to be short, have been introduced
to account for the heavy-tailed distributions observed
in the waiting time and response time distributions
\cite{alb,darwin,vazquez,caldarelli,cobham,grinstein,masuda}.
The priority queue model is grounded on the assumption
that the human dynamics is the result of an inherent decision-making
process of the individual, with implicit priorities assigned
for every tasks in his/her task queue, according to which
he/she decides which task to execute next.

To be specific, the priority queue model by Barab\'asi \cite{alb} 
consists of a single fixed-length queue, filled with tasks
each of which is assigned a priority value drawn randomly
when it enters into the queue. Every step the task with the highest
priority is executed and is replaced by a new task with random priority value.
Upon execution, the waiting time $\tau$,
that is, how long the task has sat (waited) on the queue, is measured. 
Waiting time distribution $P(\tau)$ of the Barab\'asi model has been shown to 
exhibit a power-law tail for large $\tau$ as  
\begin{equation} P(\tau)\sim \tau^{-\alpha},  \end{equation}
with the exponent $\alpha=1$ \cite{alb,vazquez}, 
conforming to the behaviors observed for the e-mail, library loan, 
and website visitation records \cite{alb,pre,goh-alb,amaral}.
Besides the human dynamics, however, due to the extremal nature of 
its dynamics the priority queue model would bear implications also
to disparate problems in extremal dynamics, such as the Bak-Sneppen model 
for biological evolution \cite{bs} and invasion percolation \cite{caldarelli}.

Barab\'asi model purposefully simplified many aspects of potential importance 
in realistic human dynamics, serving as a starting framework on which
various detailed factors can be embedded \cite{memory,blanchard,web,activity,ov}.
One important factor that was not been accounted for is the human interaction.
In the modern society, human engages in a large array of interactions 
with other individuals in various modes. As a result, typical activity 
of a person is not an outcome of completely autonomous decisions, but 
of delicate compromises and balanced conflicts between often competing
priorities. The impact of such a human interaction on the patterns
of human dynamics, the waiting time distributions in particular,
has been addressed recently by Oliveira and Vazquez (OV) \cite{ov}. 
They introduced a minimal model consisting of two interacting priority queues 
with interacting (I) and non-interacting (O) tasks. 
The human interaction is taken account for in a way that 
the $I$-task is executed only when both of the individual choose to 
execute them, that is, an AND-type protocol for the execution of $I$-task.
Through this model they showed that the power-law waiting time 
distribution still persists against the introduction of human interaction,
but it has an effect that the power-law exponent $\alpha$ of $P(\tau)$ 
can take numerable values other than $1$ depending on the queue length.

Yet, the effect of human interactions for a system of more than
two queues, or the queue {\em network} in general, has not been fully addressed.
Such a question should be highly meaningful given the increasingly 
active engagement in various social networking of individuals 
forming complex network structures \cite{rmp,guido-book,social}. 
In this regard, here we study the dynamics of priority-queue network
by generalizing the OV model.
Specifically we focus on the scalability of the interaction protocols
and the waiting time distribution under various model settings 
such as the number of queues (network size) and interaction topology.

{\em Priority-queue network}---
To construct a priority-queue network of $N$ queues, we follow
the OV model to divide the tasks into two classes; $I$- and $O$-tasks.
A queue has one $I$-task for each neighbor in the network, in
addition to an $O$-task. Thus a queue node $i$ with degree $k_i$ 
(degree is the number of links connected to the node)
has a queue with fixed length $L_i=k_i+1$. We denote the $I$-task
of the node $i$ paired with the node $j$ as $I_{ij}$,
and the $O$-task of the node $i$ as $O_i$.
Given the network configuration and the queue discipline
such as the interaction ({\it e.g.}, AND- or OR-type) and update protocols 
({\it e.g.}, parallel or sequential), 
a priority-queue network is specified.
Initially each task is given a priority value drawn from a uniform
distribution in $[0,1)$.
Then each step, each node chooses its highest priority task.
The execution of the selected tasks is determined by 
the queue discipline: In the AND-type protocol, random sequential update
case (the OV model), for example, we choose a random node, say $i$.
If the highest priority task of $i$ is $O_i$, then it is
executed. If it is an $I$-task, say $I_{ij}$, it
is executed only if $I_{ji}$ is also the highest priority
task of the conjugate node $j$. Otherwise, node $i$ executes $O_i$ instead.
The waiting times of the executed tasks are recorded, 
and the executed tasks are replaced with new
tasks each with a random priority value in uniform $[0,1)$.
$N$ such updates constitute a Monte Carlo step (MCS),
which is the time unit of waiting time measurement.

{\em OV model on networks}---
We first consider the generalization of the OV model for $N>2$ queues. 
We consider the model on two representative network configurations, 
the star graph and fully-connected network, for various $N$.
The resulting waiting time dynamics reveal
an important phenomenon, the dynamic freezing due to priority conflict. 
The priority conflict occurs when a node $i$ has $I_{ij}$ as highest priority task,
but the node $j$ has another, say $I_{jk}$, as its highest priority,
in conflict with each other.

The star topology is less vulnerable to such a dynamic freezing
since leaf-nodes can resolve it, primarily by updating priority of
the $O$-task repeatedly. As a result, we have a power-law
decaying $P(\tau)$ (Fig.~1a-b). 
The power-law exponent $\alpha$ is found to be independent of the 
network size $N$;
for the $I$-tasks $\alpha_I\approx2$, and for the $O$-tasks
$\alpha_O\approx3$ irrespective of being hub or leaf nodes.
This result is consistent with the OV model with $L=2$,
so on top of star graph, it behaves essentially
the same as in the binary OV model.

On the other hand, the dynamics is quite different in loopy networks 
such as the fully-connected networks, which are highly susceptible 
to conflicts that cannot be resolved readily.
As a consequence, the number 
of executed $I$-tasks, $\eta(t)$, decays rapidly in time,
either algebraically for small $N$, or exponentially 
for large $N\gtrsim 10$ (Fig.~1c-d), and eventually
the dynamics gets frozen, with the time scale decreasing with $N$. 
In real social networks, we have strong empirical evidences 
of high propensity of transitive triad relations (cyclic 
interactions between three individuals) \cite{social,clustering}
and clique (fully-connected subgraph) structure \cite{social,clique},
so the AND-type interaction protocol would strongly suffer 
the dynamic freezing, rendering itself unrealistic
towards realistic modeling of the network effects in human dynamics.

\begin{figure}
\centerline{\epsfxsize=9cm \epsfbox{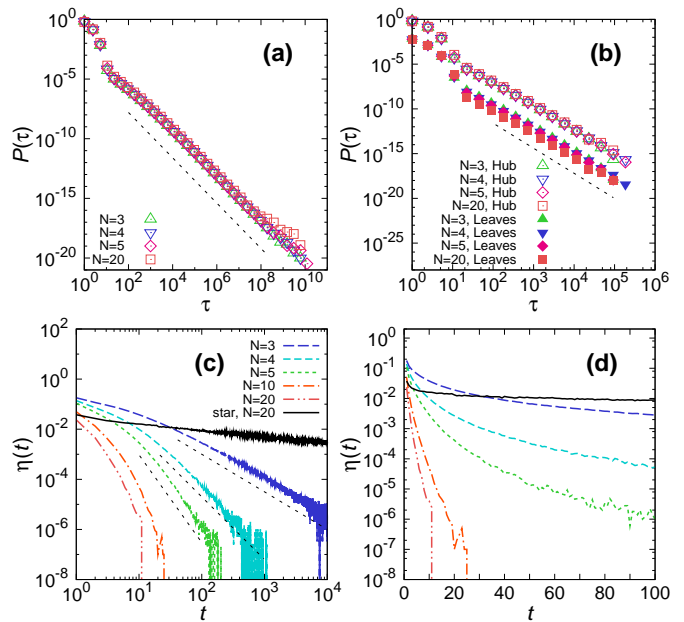}}
\caption{(Color Online) {\bf (a--b)}
The waiting time distribution $P(\tau)$ of the OV model
on star networks for $I$-tasks (a) and for $O$-tasks (b), 
with various $N=3,4,5,20$.
Both $P(\tau)$ decay with an asymptotic power law with $N$-independent exponent,
$\alpha_I\approx1.9$ (a) and $\alpha_O\approx2.8$ (b), 
indicated respectively with dotted lines.
Deviations from the power law for large $\tau$ for large $N$ are
due to the finite simulation time ($10^{10}$ steps).
{\bf (c--d)} The double logarithmic (c) and semi-logarithmic (d) plots of 
the number of executed $I$-tasks, $\eta(t)$, for the OV model
on fully-connected networks starting from random initial priority
assignments versus time.
Different dotted line patterns are used for different network size $N$
(see legend).
For small $N$, $\eta(t)$ decays algebraically with $N$-dependent exponent (c)
while it decays exponentially for large $N\gtrsim10$~(d).
Indicated slopes of dotted line are 1.5, 2.5, and 3.5, 
from right to left, drawn for the eye.
Also shown is $\eta(t)$ for $N=20$ with the star graph topology (black solid)
for comparison.
}
\end{figure}

\begin{figure}
\centerline{\epsfxsize=9cm \epsfbox{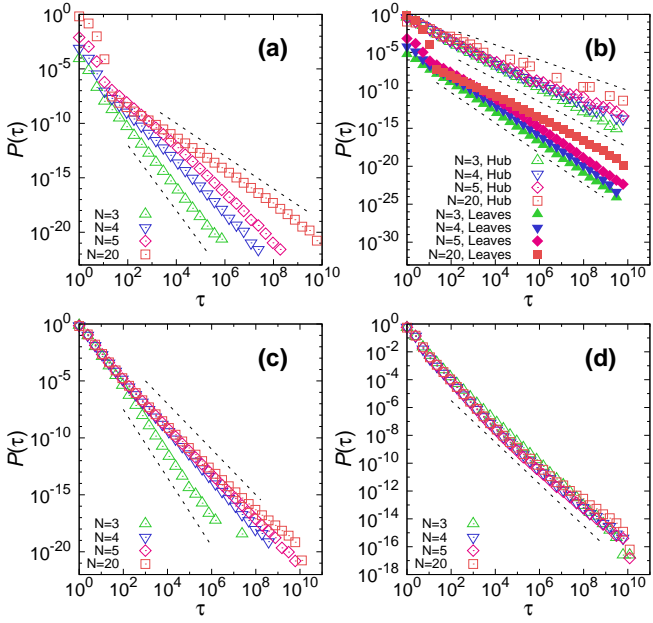}}
\caption{(Color online) 
{\bf (a--b)} The waiting time distribution $P(\tau)$ 
of the OR model on star topology with $N=3,4,5,20$.
(a) $P(\tau)$ for $I$-tasks decay as power laws asymptotically
and the power-law exponent decreases as $N$ from $\alpha_I\approx3$ for $N=3$
to $\alpha_I\approx1.5$ for $N=20$.
(b) $P(\tau)$ for $O$-tasks show distinct behaviors 
between the hub node (open symbols) and leaf nodes (full symbols).
For the hub node, the power-law exponent varies from
$\alpha_{O,hub}\approx1.5$ for $N=3$ to $\alpha\approx1$ for $N=20$.
For the leaf nodes, $P(\tau)$ decays faster, with exponents
ranging from $\alpha_{O,leaf}\approx 2$ for $N=3$ to
$\alpha_{O,leaf}\approx1.5$ for $N=20$.
For $I$-task and leaf nodes' $O$-task, we shifted $P(\tau)$
curve vertically to enhance visibility.
{\bf (c--d)} $P(\tau)$ of the OR model on fully-connected topology.
(c) $P(\tau)$ for $I$-tasks decay with asymptotic powers,
with exponents decreasing with $N$ from $\alpha_I\approx3$ for $N=3$
to $\alpha_I\approx2$ for $N=20$.
(d) $P(\tau)$ for $O$-tasks follow the power-law decay
with $N$-insensitive exponent $\alpha_O\approx 1.5$.
Deviations for large $\tau$ are due to finite simulation time.
All quoted slopes are indicated with dotted lines drawn for the eye.
}
\end{figure}

{\em Priority-queue network with OR-type protocol: The OR model}---
Not all human $I$-tasks should follow the AND-type
protocol. As an alternative, an OR-type protocol would
be more reasonable for the tasks
which require simultaneous actions of two or more individuals
but the action can be initiated primarily by either of them,
such as the phone call conversation \cite{phone2} 
and the instant messaging \cite{msn}.
For such class of $I$-tasks, the potential priority conflict
can be instantly overridden; 
we normally just answer the incoming phone call, for example.
To model such situations, we introduce the priority-queue
network with the OR-type interaction protocol by
modifying the OV model as follows:
$a)$ Each step we choose a random node, say node $i$.
$b)$ If its highest priority task is an $I$-task, say $I_{ij}$,
the two tasks $I_{ij}$ and $I_{ji}$ are executed regardless of 
the priority value of $I_{ji}$;
If $O_i$ is the highest priority task, only that is executed. 
$c)$ Priorities of all the executed tasks are randomly reassigned.
We refer this model to as the OR model hereafter.

For the OR model, $P(\tau)$ still exhibit power-law tails
for both star and fully-connected network topology,
yet the power-law exponent $\alpha$ depends on 
the network size $N$ as well as the network topology in a diverse way.
First, in the star topology, $\alpha$ decreases as $N$ increases:
For $I$-tasks, it exhibits values from $\alpha_I\approx3$ for $N=3$ 
to $\alpha_I\approx1.5$ for $N=20$ (Fig.~2a);
For $O$-tasks, the exponent exhibits distinct values for the hub
and leaf nodes, changing from $\alpha_{O,hub}\approx1.5$ for $N=3$ to
$\alpha_{O,hub}\approx1.0$ for $N=20$ for the hub node,
whereas for the leaf nodes it changes from $\alpha_{O,leaf}\approx2.0$ 
for $N=3$ to $\alpha_{O,leaf}\approx1.5$ for $N=20$ (Fig.~2b).
Moreover, for the hub-node, the mean waiting time $\langle\tau\rangle_{O,hub}$
of $O$-tasks diverges with the exponent $\alpha<2$, similarly to
the Poisson queue placed on the hubs in scale-free 
networks \cite{hklee}. For other tasks, $P(\tau)$ with
$\alpha<2$ for large $N$ is accompanies by the peak at $\tau=1$,
rendering the average waiting time finite.

In the fully-connected topology, the power-law exponent $\alpha$ 
weakly depends on $N$. For the $I$-tasks,
it decreases with $N$ from $\alpha_I\approx3$ for $N=3$ 
to $\alpha_I\approx2$ for $N=20$ (Fig.~2c).
For the $O$-tasks, on the other hand, 
$\alpha_O$ is rather stable against $N$ as $\alpha_O\approx1.5$ (Fig.~2d).
This result implies that on the fully-connected networks,
$I$-tasks are executed with finite mean waiting times
while $O$-tasks on average have to wait on the queue 
infinitely long to be executed.
Taken together, the OR model on networks implicates
the importance of not only the overall
network structure but also individual node's topological position
on determining the dynamics of networking priority-queue nodes.

\begin{figure}
\centerline{\epsfxsize=9cm \epsfbox{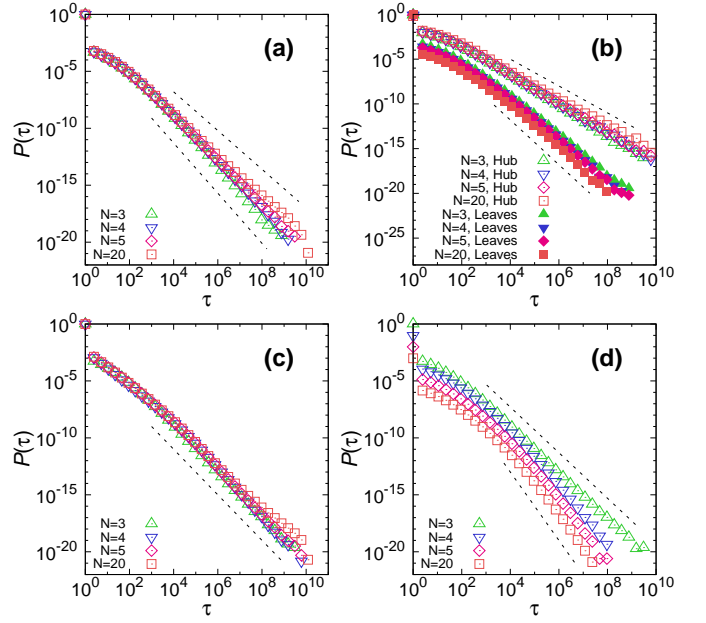}}
\caption{(Color online) 
The waiting time distribution $P(\tau)$ of the OR model with parallel updates.
{\bf (a--b)} $P(\tau)$ of the parallel OR model on star topology 
for the $I$-tasks (a) and the $O$-tasks (b).
For $I$-tasks, the power-law decay of $P(\tau)$ exhibits
a weak $N$-dependence, as the exponent varies from $\alpha_I\approx2.2$
for $N=3$ to $\alpha_I\approx1.7$ for $N=20$.
For $O$-tasks, hub and leaf nodes exhibit different $N$-insensitive 
power laws with $\alpha_{I,hub}\approx 1.5$ for the hub
and $\alpha_{I,leaf}\approx 2.5$ for the leaf nodes.
{\bf (c--d)} $P(\tau)$ of the parallel OR model on fully-connected topology.
For $I$-tasks the power-law exponent is found to be insensitive 
to $N$ as $\alpha_I\approx2$ (c).
For $O$-tasks, however, the asymptotic power-law exponent 
increases with $N$, from $\alpha_O\approx2$ for $N=3$ 
to $\alpha_O\approx3$ for $N=20$. In (d), the $P(\tau)$
curves for different $N$ are shifted vertically to enhance comprehensibility.
All quoted slopes are indicated with dotted lines drawn for the eye.
}
\end{figure}

{\em Priority-queue network with parallel updates}---
Update rule in discrete time dynamic models has been
known to affect the dynamics considerably \cite{asep}.
Thus it is informative to study the effect of update rule
in priority-queue models.
To this end, we consider the parallel update rule
by modifying the OR model as follows:
$a)$ Each step, each node chooses its highest priority task.
$b)$ We sort all chosen tasks by the priority values,
and execute them in order of priority, while each node 
can execute at most one task each step.
That is, if the priority of $I_{ij}$ is higher than that of $I_{jk}$,
than the node $j$ executes the task $I_{ji}$ upon request from $i$
before $I_{jk}$, which subsequently cannot be executed in this step.
$c)$ All the executed tasks are assigned 
new random priorities, completing a MCS.
Priority conflict may occur at the step $a)$, but 
it is not as problematic as in the AND-type protocol case even
in the fully-connected topology, due to the partial resolution in $b)$.
We found, however, that it is strong enough
to affect the waiting dynamics of the priority-queue network:
it can reshape $P(\tau)$ in a significant way, 
because the tasks in the tail $(\tau\neq1)$
are particularly strongly affected.

The waiting time distribution $P(\tau)$ of the OR model with
parallel update unanimously exhibits a probability
weight strongly concentrated at $\tau=1$ followed
by an asymptotic algebraic tail (Fig.~3).
On the star graph topology, the decay exponent $\alpha$
shows less variation compared to the sequential update case:
It changes from $\alpha_I\approx2.2$ for $N=3$ to 
$\alpha_I\approx1.7$ for $N=20$ for the $I$-tasks (Fig.~3a).
For the $O$-tasks, it is insensitive to $N$, yet
exhibits different values for the hub and leaf nodes,
as $\alpha_{O,hub}\approx1.5$ and $\alpha_{O,leaf}\approx2.5$, 
respectively (Fig.~3b), decaying faster than the sequential case.
For the fully-connected networks, the power-law decay exponent
for the $I$-tasks is $\alpha_I\approx2$, insensitive to $N$ (Fig.~3c),
while for the $O$-tasks, it even increases with $N$, 
from $\alpha_O\approx 2$ for $N=3$ to $\alpha_O\approx3$ for $N=20$ (Fig.~3d).
The presence of strong peak at $\tau=1$ renders the mean waiting
time of all tasks finite, which is due to the partial
resolution rule we implemented in the model.
These results demonstrate clearly that the update
rule in the priority queue network strongly affect the overall
dynamics in a nontrivial way.

{\em Summary and Discussion}---
In this paper, we have generalized the binary interacting priority
queue model, the OV model, into a queue network, showing that
the OV model is not easily generalizable onto loopy networks
as the dynamics gets frozen due to substantial priority conflicts.
We then introduced a modified model, the OR model,
which can be put on top of any network topology. 
It is shown that OR model exhibits power-law decaying waiting
time distribution $P(\tau)$, Eq.~(1), yet with diverse values 
of the exponent $\alpha$ depending on the global network topology, 
local position of queue nodes on the network,
as well as the queue discipline such as the update rule. 

The fundamental factor driving the diverse behaviors of
the interacting priority queue models is the existence of priority conflicts.
Different global network topology and distinct network position of nodes
impose different degree of conflicts and the resolution thereof.
In the perspective of human dynamics modeling, its introduction
seems reasonable, for such conflicts are more of a rule than an exception 
in daily decision making of modern human life. 
A crucial question remains at this point. The dependence of
waiting time dynamics of a priority queue node (a human)
on the global or local network topology that we found in this work
has never been gauged from the empirical data yet.
Candidate datasets for this end would be the mobile phone
data \cite{phone2} or the instant messaging data \cite{msn},
for which power-law-like waiting time distributions have been reported.
The waiting time $\tau$ in this work is measured for each task, 
that is for each link, differently from what have been previously 
measured from data in Refs.~\cite{phone2,msn}, so a direct comparison cannot be made.
Appropriate measurements with these datasets would reveal the relevance
of the priority-queue network models studied in this work and the role
of interactions in human dynamics in general.

\begin{acknowledgments}
This work was supported by the Korea Research Foundation Grant funded by 
the Korean Government (MOEHRD) (KRF-2007-331-C00111).
\end{acknowledgments}

\end{document}